\documentclass[referee]{raa}           
\usepackage{graphicx,times}           
\usepackage{natbib}
\usepackage{adjustbox}
\usepackage{blindtext}
\usepackage{amssymb,amsmath}
\usepackage[a4paper, total={6in, 8in}]{geometry}
\bibpunct{(}{)}{;}{a}{}{,}
\usepackage[pagebackref=true]{hyperref}
\begin{document}
\title{Classifying Galaxy Morphologies with Few-Shot Learning}

\volnopage{Vol.0 (20xx) No.0, 000--000}     
\setcounter{page}{1}          
\author{Zhirui Zhang\inst{1}\thanks{First author} 
           Zhiqiang Zou\inst{1,2\star} 
          Nan Li\inst{3,4}\thanks{Corresponding author, nan.li@nao.cas.cn}
          \and
          Yanli Chen \inst{1,2} 
          }
\institute{
         College of Computer, Nanjing University of Posts and Telecommunications, Nanjing, China 
         \and
         Jiangsu Key Laboratory of Big Data Security and Intelligent Processing, Nanjing, Jiangsu, China
         \and
         Key Laboratory of Optical Astronomy, National Astronomical Observatories, Chinese Academy of Sciences, Beijing, People’s Republic of China 
         \and
         University of Chinese Academy of Science, Beijing, People's Republic of China  \\     
         {\small Received 20xx month day; accepted 20xx month day}
    }

\abstract{ The taxonomy of galaxy morphology is critical in astrophysics as the morphological properties are powerful tracers of galaxy evolution. With the upcoming Large-scale Imaging Surveys, billions of galaxy images challenge astronomers to accomplish the classification task by applying traditional methods or human inspection. Consequently, machine learning, in particular supervised deep learning, has been widely employed to classify galaxy morphologies recently due to its exceptional automation, efficiency, and accuracy. However, supervised deep learning requires extensive training sets, which causes considerable workloads; also, the results are strongly dependent on the characteristics of training sets, which leads to biased outcomes potentially. In this study, we attempt \textit{Few-shot Learning} to bypass the two issues. Our research adopts the dataset from Galaxy Zoo Challenge Project on Kaggle, and we divide it into five categories according to the corresponding truth table. By classifying the above dataset utilizing few-shot learning based on Siamese Networks and supervised deep learning based on AlexNet, VGG\_16, and ResNet\_50 trained with different volumes of training sets separately, we find that few-shot learning achieves the highest accuracy in most cases, and the most significant improvement is $21\%$ compared to AlexNet when the training sets contain 1000 images. In addition, to guarantee the accuracy is no less than $90\%$, few-shot learning needs $\sim$6300 images for training, while ResNet\_50 requires $\sim$13000 images. Considering the advantages stated above, foreseeably, few-shot learning is suitable for the taxonomy of galaxy morphology and even for identifying rare astrophysical objects, despite limited training sets consisting of observational data only. 
}

\authorrunning{Zhirui Zhang, Zhiqiang Zou, Nan Li, Yanli Chen}  
\titlerunning{Classifying Galaxy Morphologies with Few-Shot Learning}  
\maketitle

\section{Introduction}       
\label{Introduction}
Galaxy morphology is considered a powerful tracer to infer the formation history and evolution of galaxies, and it is correlated with many physical properties of galaxies, such as stellar populations, mass distribution, and dynamics. Hubble invented a morphological classification scheme for galaxies \citep{hubble1926} and pioneeringly revealed the correlation between galaxy evolutionary stages and their appearance in optical bands. Hubble sequence principally includes early-type galaxies (ETGs) and late-type galaxies (LTGs); ETGs mostly contain older stellar populations and have few spiral structures, while LTGs hold younger stellar populations and present spiral arms-like features usually. The above correlation has been studied widely and deeply in the past decades with increasing observational data of galaxies. Predictably, relevant investigations will be significantly advanced with enormous data from the upcoming large-scale imaging surveys, such as LSST \footnote{\url{https://www.lsst.org}}, Euclid \footnote{\url{https://www.euclid-ec.org/} } and CSST \footnote{\url{http://www.bao.ac.cn/csst/}}.

Galaxy morphology classification is started with visual assessment \citep{1959,de1964luminosity,sandage1961,fukugita2007,nair2010catalog,baillard2011} and has lasted for decades as the mainstream approach in the field. In the 21st century, the volume and complexity of astronomical imaging data have increased significantly with the capability of the new observational instruments, such as the Sloan Digital Sky Survey \footnote{\url{https://www.sdss.org/}} (SDSS) and the Hubble Space Telescopes \footnote{\url{https://hubblesite.org/}} (HST). To make the classification more efficient and accurate, astronomers developed non-parametric methods to extract morphological features of galaxies, such as the concentration-asymmetry-smoothness/clumpiness (CAS) system, the Gini coefficient, and the M20 parameter \citep{conselice2003, abraham2003new, lotz2004new, law2007physical}. Sets of evidence demonstrate the success of utilizing these approaches to represent galaxy morphologies, outperforming traditional human inspection since they eliminate subjective biases. However, encountering hundreds of millions or even billions of galaxy images from future surveys, the performance of the above CPU-based algorithms is inefficient. Hence, more effective techniques for Galaxy morphology classification in an automated manner, e.g., machine learning, are necessary.

Machine learning algorithms have been widely used to classify galaxy morphology in the past years, for instance, Artificial Neural Network \citep{naim1995}, NN $+$ local weighted regression \citep{dela2004}, Random Forest \citep{gauci2010}, linear discriminant analysis (LDA, \citep{ferrari2015}). Recently, deep learning has become more and more popular for classifying galaxy morphology \citep{lukic2019, zhu2019, cheng2020} as its success has been proved adequately in industries, especially for pattern recognition, image description, and anomalies detection. Most cases for classifying galaxy morphologies are based on supervised deep learning due to its high efficiency and accuracy. Successful cases include generating the catalogs of galaxy morphologies for SDSS, the Dark Energy Survey \footnote{\url{https://www.darkenergysurvey.org/}} (DES), and the Hyper Suprime-Cam \footnote{\url{https://www.naoj.org/Projects/HSC/}} (HSC) Surveys \citep{dieleman2015,flaugher2015dark,aihara2018hyper}. However, the results of supervised deep learning are strongly dependent on the volume and characteristics of the training set. Firstly, requiring a large volume of data for training is determined by the complexity of the convolution neural networks, which typically comprise millions of trainable parameters. Hence, to make the training procedure converge correctly, one has to provide data points with a comparable amount to the number of parameters of the CNN. Secondly, the best trained CNN model reflects the properties of the feature space covered by the training set. Thus, if the training set (simulated or selected by astronomers) is biased considerably from the real Universe, supervised deep learning may give biased results consequently.  Unsupervised learning has been adopted to avoid those disadvantages, but the corresponding classification accuracy is $\sim 10\%$ worse than that of supervised manners \citep{cheng2020,cheng2021}. 

In this study, we attempt few-shot learning \citep{wang2019} to classifying galaxy morphologies by proposing a model named SC-Net inspired by Convolutional Neural Networks (CNNs) and the siamese network model \citep{chopra2005}. Concisely speaking, our method pairs images and compares the metrics between features of input images, which expands the sample size of the training set compared to feeding images directly into the CNN. Furthermore, the region of feature space covered by the training set can be enlarged more effectively by involving rare objects and pairing them with other objects. Thus, in principle, SC-Net model simultaneously improves the two drawbacks of traditional supervised deep learning. To quantify the improvements, we designed an experiment with adopting galaxy images from the Galaxy Zoo Data Challenge Project on Kaggle \footnote{\url{https://www.kaggle.com/c/galaxy-zoo-the-galaxy-challenge}} based on Galaxy Zoo 2 Project \citep{willett2013}, then compared the classification results to those with AlexNet \citep{krizhevsky2012}, VGG\_16 \citep{simonyan2014}, and ResNet\_50 \citep{he2015}. The outcomes show that our method achieves the highest accuracy in most cases and requests the most miniature training set to satisfy a given accuracy threshold (see Sect.\ref{Results} for more details). Therefore, foreseeably, SC-Net model is suitable for classifying galaxy morphology and even for identifying rare astrophysical objects in the upcoming gigantic astronomical datasets. The code and dataset used in this study are publicly available online \footnote{\url{https://github.com/JavaBirda/Galaxy-Morphologies-}}.
   
The paper is organized as follows: Sect. \ref{Datasets} introduces the datasets and data enhancement. Deep learning models, including CNNs and siamese Network, are described in Sect. \ref{Methodology}. Sect. \ref{Experiments} presents the experimental process of this study. Results of this work are analyzed and summarized in Sect. \ref{Results}. Finally, we draw discussion and conclusions in Sect. \ref{Discussion and Conclusion}.

\section{Datasets}
\label{Datasets}

The Sloan Digital Sky Survey captured around one million galaxy images. To classify the galaxy morphology, the Galaxy Zoo Project was launched \citep{lintott2008galaxy}, which is a crowd-sourced astronomy project inviting people to assist in the morphological classification of large numbers of galaxies. The dataset we adopted is one of the legacies of the galaxy zoo project, and it is publicly available online for the Galaxy-zoo Data Challenge Project. 
 
The dataset provides 28793 galaxy morphology images with middle filters available in SDSS (g, r, and i) and a truth table including 37 parameters for describing the morphology of each galaxy. The 37 parameters are between 0 and 1 to represent the probability distribution of galaxy morphology in 11 tasks and 37 responses \citep{willett2013}. Higher response values indicate that more people recognize the corresponding features in the images of given galaxies. The catalog is further debiased to match a more consistent question tree of galaxy morphology classification \citep{hart2016galaxy}.

To simply the classification problem, we reorganize 28793 images into five categories: completely round smooth, in-between smooth, cigar-shaped smooth, edge-on, and spiral, according to the 37 parameters in the truth table. The filtering method refers to the threshold discrimination criteria in \cite{zhu2019}. For instance,  when selecting the completely round smooth, values are chosen as follows: $f_{smooth}$ more than 0.469, $f_{complete,round}$ more than 0.50, as shown in Tab. \ref{tab1}.

We then build six training sets with different numbers of images to test the dependence of the performance of classification algorithms on the volume of training sets, details of the training sets are shown in Tab. \ref{tab2}. In Sect. \ref{Experiments}, we will train all the deep learning models with 28793, 20000, 15000, 10000, 5000, and 1000 images, respectively, and compare their performances thoroughly.

\begin{table}[]
\caption{The classification of 28793 samples. The first column is the name of categories, the second column is the threshold selection for filtering data, and the third column is the size of each category.}\label{tab1}
    \centering
\begin{tabular}{lll}
\hline
Class-name              & Thresholds                & Number \\ \hline
Completely round smooth & $f_{smooth} \geq 0.469        $   & 8436   \\ \hline
                        & $f_{completely round} \geq 0.50 $ &        \\
In-between smooth       & $f_{smooth} \geq 0.469        $   & 8069   \\
                        & $f_{in-between} \geq 0.50     $   &        \\ \hline
Cigar-shaped smooth     & $f_{smooth} \geq 0.469        $   & 579    \\
                        & $f_{cigar-shaped} \geq 0.50   $   &        \\ \hline
Edge-on                 & $f_{features/disk} \geq 0.430 $   & 3903   \\
                        & $f_{edge-on,yes} \geq 0.602   $   &        \\ \hline
Spiral                  & $f_{edge-on,no} \geq 0.715    $   & 7806   \\
                        & $f_{spiral,yes} \geq 0.619    $   &        \\ \hline
Total                   &                              & 28793   \\
                        \hline

\end{tabular}
\end{table}
 
\begin{table}[]
\caption{The amount of data in each category under different data sizes. The first column represents the size of the data, and the next five columns represent the size of the data of completely round smooth, in-between smooth, cigar shaped smooth, edge-on and spiral, and the above five categories are represented by 0, 1, 2, 3 and 4, respectively.}\label{tab2}
    \centering
\begin{tabular}{cccccc}\hline
Total size & 0    & 1    & 2   & 3    & 4    \\ \hline
28793      & 8436 & 8069 & 579 & 3903 & 7806 \\ \hline
20000      & 5237 & 5071 & 371 & 2420 & 4901 \\ \hline
15000      & 3887 & 3785 & 283 & 1811 & 3734 \\ \hline
10000      & 2601 & 2516 & 197 & 1219 & 2467 \\ \hline
5000       & 1298 & 1261 & 87  & 607  & 1247 \\ \hline
1000       & 269  & 241  & 19  & 121  & 250  \\ \hline
\end{tabular}
\end{table}

\section{Methodology}
\label{Methodology}

The few-shot learning proposed in this study is based on a model named SC-Net, including a CNN and a siamese network. We use CNNs to extract features, and then train the model according to the idea of the siamese network for classifying galaxy morphologies. Explicitly, the CNNs section introduces the feature extraction process and several traditional CNNs \citep{krizhevsky2012,simonyan2014,he2015,lecun1998} for classification; the siamese network section describes the few-shot learning method and the structure of the siamese network. 
\subsection{Convolution neural networks}
\label{Convolution neural network}
CNN is a feed forward neural networks which includes convolutional computation and deep structure, and is one of the representative algorithms of deep learning. CNN is essentially input-to-output mappings that learns mapping relationships between inputs and outputs without requiring any precise mathematical expressions so that CNN has been widely used in the field of computer vision in recent years.
 
The schematic of image feature extraction with CNN mainly consists of the following three main layers: convolution layer, pooling layer and fully connected layer. The convolution layer for feature extraction of the image is built by dot multiplication operation of the image and convolution kernel. Each pixel of the image and the weight of the convolutional kernel are computed through convolution layer and the calculation process is shown in Eq. (\ref{eq1})
\begin{equation}
    \label{eq1}
   a_{i,j}=f(\sum_{m=0}^{2}\sum_{n=0}^{2}w_{m,n}x_{i+m,j+n}+w_{b})
\end{equation}
where the \emph{f} function is an activation function.We usually use Rectified Linear Units (Relu) \citep{glorot2011} as the activation function defined in Eq. (\ref{eq2}). $w_{m,n}$ means the weight, $x_{i+m,j+n}$ means input data of current layer, $w_b$ represents bias, and $a_{i,j}$ describes the output data of current layer. 
\begin{equation}
  \label{eq2}
  Relu(x)=max(0,x) 
\end{equation}
Relu turns a negative value into zero. The image size obtained by the convolution operation is related to certain factors, such as the size of the convolution kernel, the convolution step size, the expansion method and the image size before convolution. The formula description of convolution operation is shown in Eq. (\ref{con_op}).
\begin{align}
\label{con_op}
W_{2}=(W_{1}-F+2P)/S+1 \\
H_{2}=(H_{1}-F+2P)/S+1   
\end{align}
where $W_1$ means the width of input data, $H_1$ means the height of input data, F represents the size of convolution kernels, P describes the padding size and S means stride, $W_2$ and $H_2$ denote the value of $W_1$ and $H_1$ after being calculated. The pooling layer is applied to reduce the image size while retaining important information. Max-pooling retains the maximum value of feature map as the resulting pixel value, while average-pooling retains the average value of feature map as the resulting pixel value. The fully-connected layer acts as a “classifier” for the entire convolutional neural network after convolution, activation function, pooling and other deep networks. The classifying results are identified by the fully-connected layer.

In the past 20 years, traditional CNN algorithms for image classification have made breakthroughs \citep{lecun1998}. AlexNet for ImageNet competition was proposed by Hinton's student Alex Krizhevsky \citep{krizhevsky2012}, which established the status of convolutional neural networks in computer vision. VGGNet \citep{simonyan2014} was proposed by the Oxford University Computer Vision Group in 2014, which has good generalization ability and can be easily migrated to other image recognition projects. Kaiming He et al. proposed the ResNet \citep{he2015} in 2015, which solves the problem of gradient explosion due to depth of model layers. 

Although the development of deep learning has made great achievements, deep learning models are strongly dependent on the size and quality of the dataset. Traditional deep learning models cannot get a better result when lacking plenty of samples. To solve this problem, some researchers introduced data augmentation methods and generate simulated samples, such as GAN \citep{2014Generative}, which alleviate the difficulty of insufficient samples to a certain extent. However, its result is not very ideal because of the deviation between the real world data and simulated samples. Therefore, new method is needed to solve this problem.

\subsection{Siamese network}
\label{Siamese network}
To solve the problem of lacking of enormous samples with high quality mentioned in Sect. \ref{Convolution neural network}, this study introduces the few-shot learning \citep{wang2019}. Few-shot learning is an application of Meta Learning \citep{schweighofer2003} in the field of supervised learning, which is mainly used to solve the problem of model training with a small number of classified samples. Few-shot learning is divided into three categories: model-based method, optimization-based method \citep{wang2019} and metric-based method.

The model-based methods aim to learn the parameters quickly over a small number of samples through the design of model structure, and directly establish the mapping function between the input value and the predicted value, such as memory-enhancing neural network \citep{santoro2016}, meta networks \citep{munkhdalai2017}. Optimization-based methods consider that ordinary gradient descent are inappropriate under few-shot scenarios, so they optimize learning strategies to complete the task of small sample classification, such as LSTM-based meta-learner model \citep{ravi2016}. The metric-based methods measure the distance between samples in the batch set and samples in the support set by using the idea of the nearest neighbor, such as siamese network \citep{koch2015}. Considering the universality and conciseness of metric distance, this study chooses metric-based method.

\begin{figure}
\centering
\includegraphics[width=\hsize]{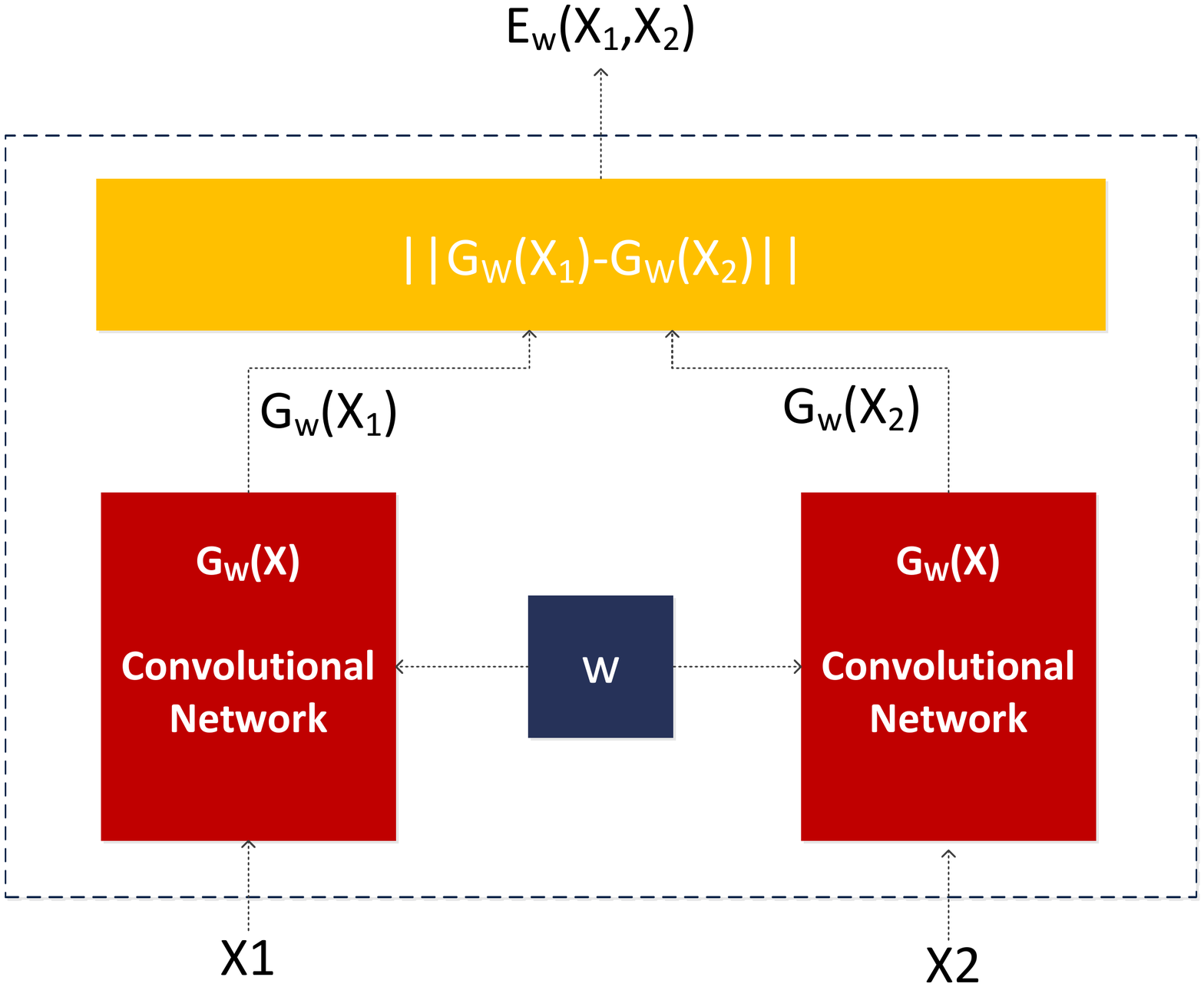} 
\caption{Siamese Architecture. The left and right input different data, and calculate the similarity between them after feature extraction.}
\label{fig1}
\end{figure}

The siamese network is a metric-based model in few-shot learning, which was first proposed in 2005 \citep{chopra2005} for face recognition. The basic idea of siamese network is to map the original image to a low-dimensional space, and then get feature vector. The distance between the feature vectors is calculated through the euclidean distance. In our study, the distance between the feature vectors from the same galaxy morphology should be as small as possible, while the distance between the feature vectors from the different galaxy morphology should be as large as possible.The framework of simaese network is shown in Fig. \ref{fig1} \citep{chopra2005}.

In siamese network, the structures of two networks on the left and right share the same weights(W). The input data, denoting ($X_1$,$X_2$,Y), are two galaxy morphology and the label that measures the difference between them. The label Y will be set to 0 when $X_1$ and $X_2$ belong to the same galaxy morphology, and it will be set to 1 when $X_1$ and $X_2$ belong to different galaxy morphology. The feature vector Gw($X_1$) and Gw($X_2$) of low-dimensional space are generated by mapping $X_1$ and $X_2$, and then their similarity is computed by Eq. (\ref{eq5}).
\begin{equation}
\label{eq5}
 E_{W}(X_{1},X_{2})=|| G_{W}(X_{1})-G_{W}(X_{2}) ||
\end{equation}

The SC-Net makes Ew($X_1$,$X_2$) as small as possible when Y=0 and makes Ew($X_1$,$X_2$) as large as possible when Y=1. Contrastive Loss \citep{hadsell2006} is selected in SC-Net as the loss function, which makes the originally similar samples are still similar after dimensionality reduction and the original dissimilar samples are still dissimilar after dimensionality reduction. The formula for the contrast loss function is shown in Eq. (\ref{eq6}).
\begin{equation}
\label{eq6} 
L(W,Y,X _{1},X_{2})=(1-Y)L_{G}(E_{W})+YL_{I}(E_{W})
\end{equation}
When the input images belong to the same galaxy morphology, the final loss function depends only on $L_G$($E_W$), and when the input images belong to different galaxy morphology, the final loss function depends on $L_I$($E_W$). $L_G$($E_W$) and $L_I$($E_W$) are defined in Eq. (\ref{eq7}) and Eq. (\ref{eq71}). The constant Q is set to the upper bound of $E_W$.
\begin{align}
\label{eq7}
L_{G}(E_{W})=\frac{2}{Q}(E_{W})^{2}\\
\label{eq71}
L_{I}(E_{W})=2Qe^{-\frac{2.77}{Q}E_{W}}
\end{align}

As so far, we can train the SC-Net model according to the architecture and loss function as described above. And then the classified results will be obtained. The advantage of this method is to fade the labels, making the network have good extension. Moreover, this approach increases the size of the data set by pairing data operation, so that deep learning network will achieve better effect with the small amount of data. For the above reasons, we adopt siamese network and put forward the SC-Net model.

\section{Experiments}
\label{Experiments}
\begin{figure}
\centering
\includegraphics[width=0.6\hsize]{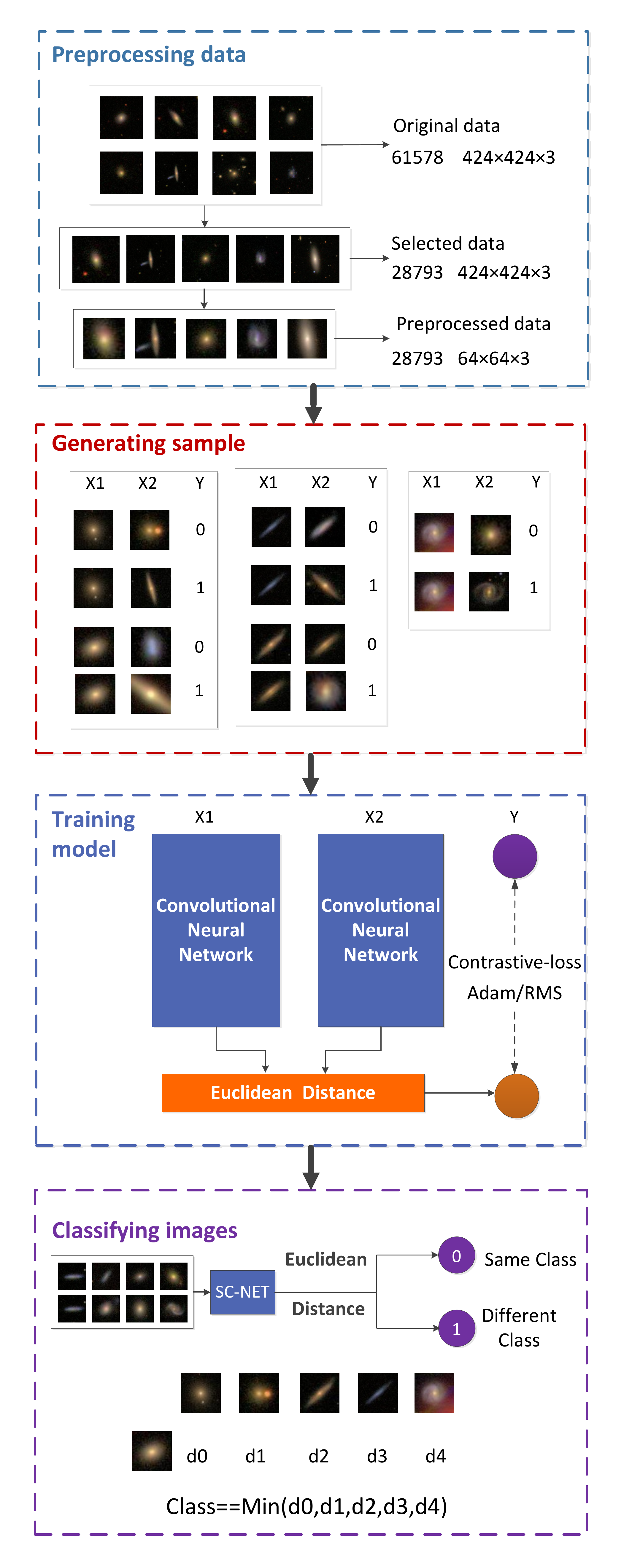}
\caption{ The workflow of SC-Net model, including data preprocessing, sample generating, model training, and image classifying.}
\label{fig2}
\end{figure}

The workflow of our SC-Net model is shown in Fig. \ref{fig2}. The whole procedure includes four stages: the first stage is to preprocess data with the method introduced in Sect. \ref{Data Pre-processing}; the second stage is to generate the training set via re-sampling or sub-sampling the preprocessed data; the third stage is to train model based on the networks described in Sect. \ref{Deep Learning Models}; the last stage is to classify the images using the trained model. Sect. \ref{Implementation Details} describes the details of the implementation of these experiments.

\subsection{Data Pre-processing}
\label{Data Pre-processing}
\begin{figure}[ht!]
\centering
\includegraphics[width=\hsize]{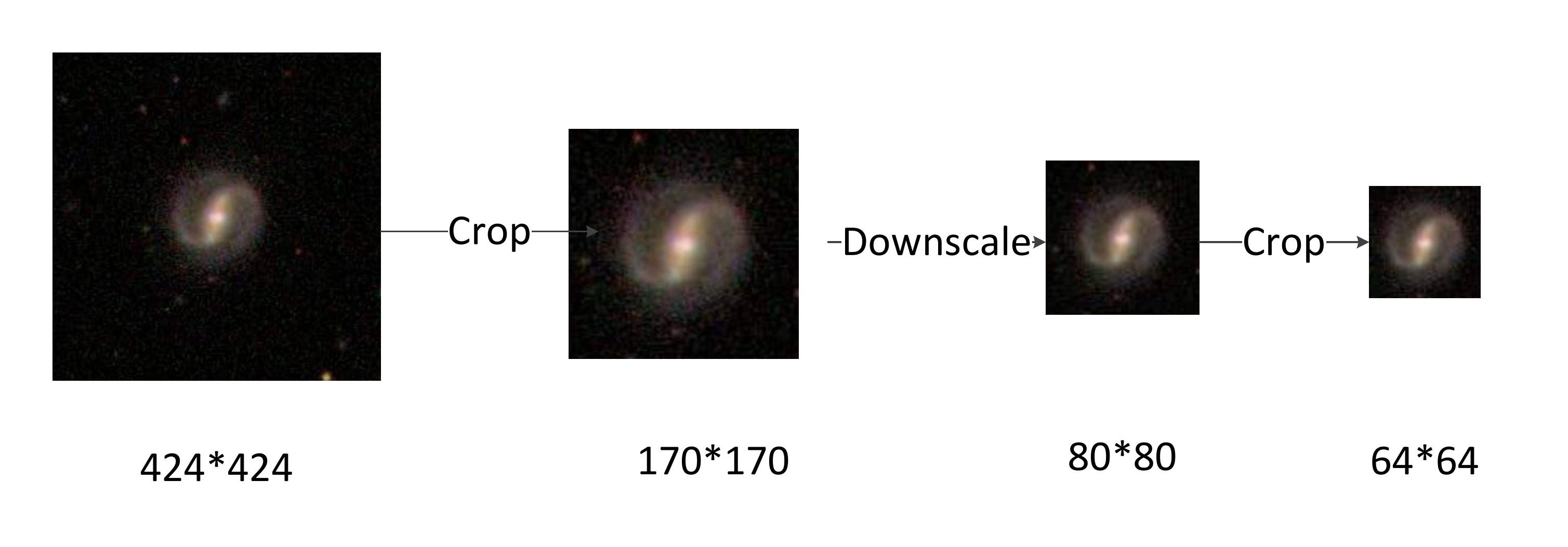}
\caption{Data processing for the image(ID 11244)  from 424×424 to 64×64.}
\label{fig3}
\end{figure}

The experiment datasets consist of 28793 images with $424\times424\times3$ pixels in size. The training time is sensitive to the size of images, and the features of galaxies are primarily concentrated at the centers of the original images. Therefore, we crop and scale the original images first, then arrange them to training sets. The workflow of data preprocessing is shown in Fig. \ref{fig3}, which is the same as shown by \citet{zhu2019}. 
We first crop the original images from $424\time424$ pixels to $170\times170$ pixels, considering the image centers as origins. Then, the images with $170\times170$ pixels are resized to $80\times80$ pixels. A last, we repeat the first step to crop the images with $80\times80$ pixels to $64\times64$ pixels. 

\begin{figure}[ht!]
\centering
\includegraphics[width=\hsize]{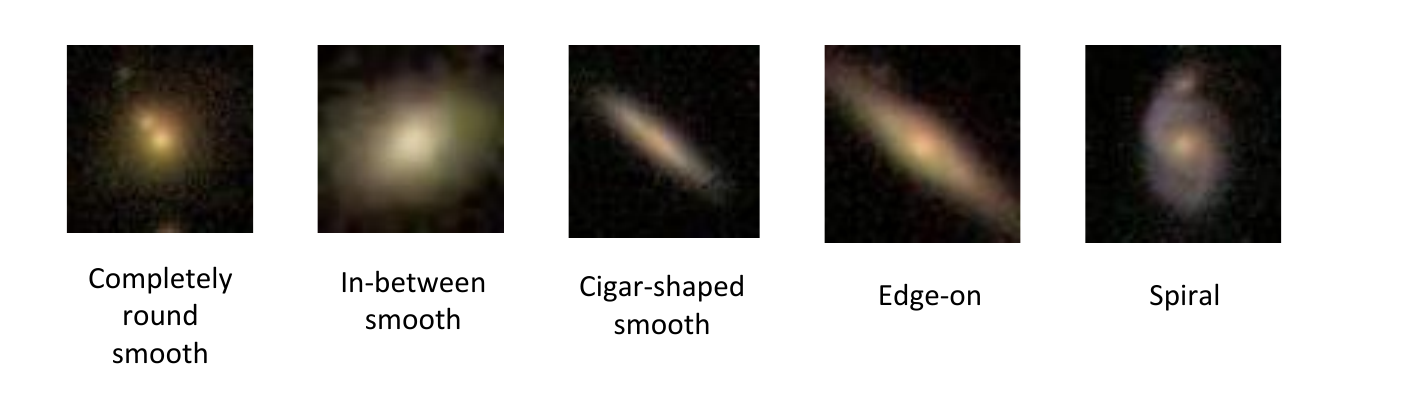}
\caption{Images sample from five categories}
\label{fig4} 
\end{figure}

As is mentioned in Sect. \ref{Datasets}, we have divided the 28793 images into five categories according to the truth table with the approach used in \citet{zhu2019} and organized six datasets to implement comparative experiments for quantifying the advantages of the SC-Net over traditional CNNs. The six datasets contain 1000, 5000, 10000, 15000, 20000, and 28793 images. The preprocessed datasets have the same organization but images with $64\times64\times3$ pixels, and examples from each category are shown in Fig. \ref{fig4}. 

The data form that the SC-Net model takes is ($X_1$, $X_2$, Y), where $X_1$ and $X_2$ represent a pair of images, and Y is the label of the correlation between $X_1$ and $X_2$. For example, $X_1$ is in category edge-on, and $X_2$ is selected in the same category, then Y was set to 0. To balance positives and negatives in training sets, every time we create a positive data point, we create a negative data point by randomly selecting an image from other categories. 

\subsection{Deep Learning Models}
\label{Deep Learning Models}
For comparison, we first build three approaches based on traditional CNNs: (1) AlexNet \citep{krizhevsky2012}, (2) VGG\_16 \citep{simonyan2014}, and (3) ResNet\_50 \citep{he2015}.   (1)AlexNet is consists of 5 convolutional layers and 3 fully connected layers. The network structure is successively conv11-96, max pool, conv5-256, maxpool, conv3-384, conv3-384, conv3-256, maxpool. (2) The network structure of VGG\_16 is constructed by modularization. The first and second modules are divided into two convolutional layers and a max-pooling layer, and the last three modules are composed of three convolutional layers and a max-pooling layer. The number of channels in the convolutional kernel increases from 64 to 512, and finally, three fully connected layers are added, with the number of neurons being 4096, 4096, and 1000 successively. There are 13 convolution layers and 3 full connection layers in total. (3) Resnet\_50 consists of a convolutional layer, 16 residual modules, and a full connection layer. The residual module has an identity block and a convolutional block composed of three convolutional layers and a shortcut. The difference lies in that the identity block ensures the consistency of input and output data. The input images of all CNN models are of $64\times64$ pixels in three channels, and the outputs are vectors of $1\times5$. The remaining parameters, such as the network hierarchy and hyperparameters, were referred to in the original papers. 

\begin{figure}
\centering
\includegraphics[width=\hsize] {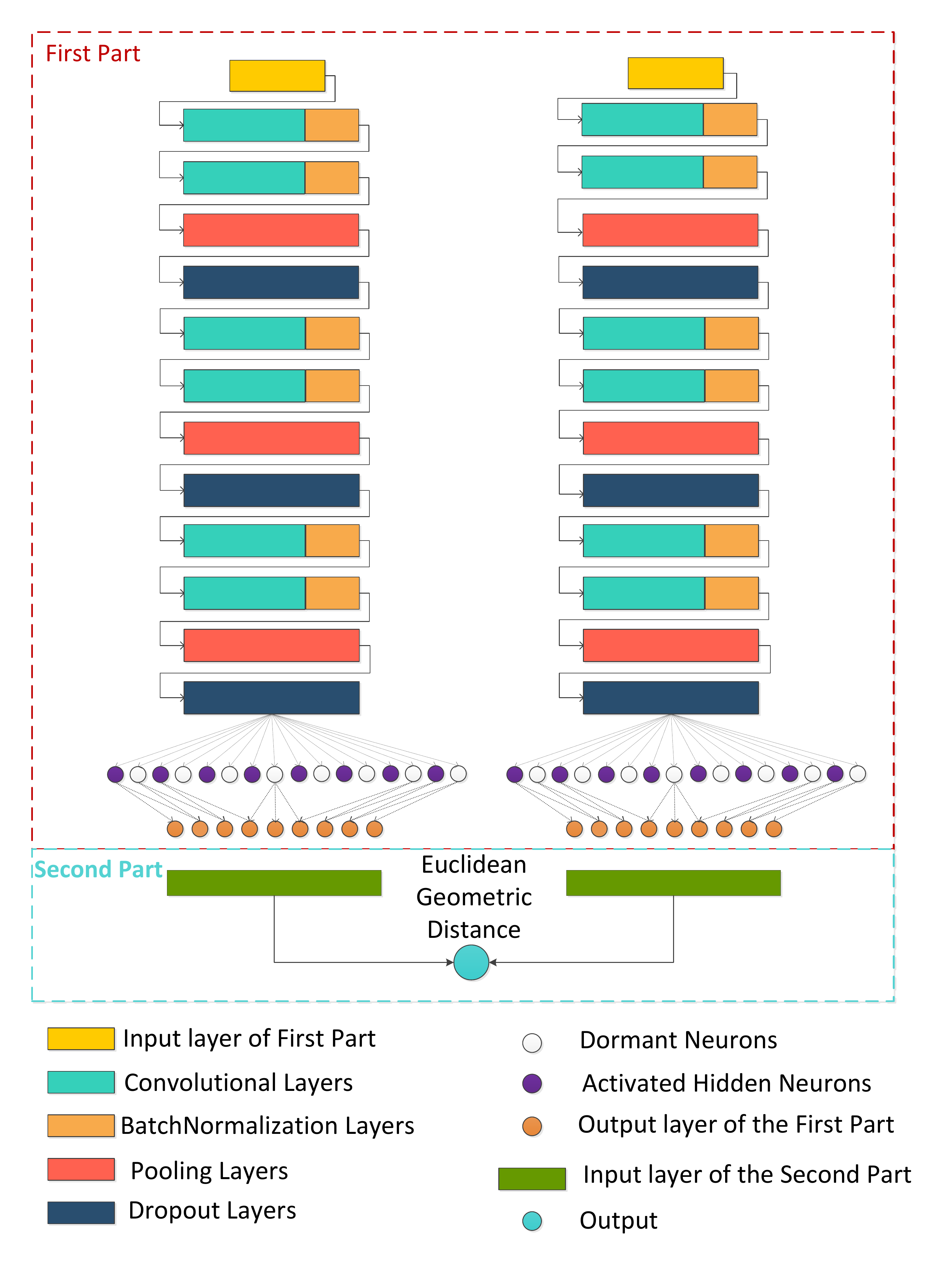}
\caption{Architecture of the SC-Net model. The meaning of each icon is explained at the bottom of the figure.}
\label{fig5}
\end{figure}

The architecture of SC-Net is shown in Fig. \ref{fig5}, which consists of two parts. The first part is for extracting features with a convolutional neural network, and the second part calculates the similarity between the feature vectors obtained from the first part. The outputs of the SC-Net are the Euclidean Distances between feature vectors of input images in the feature space, which are to be used in the classification stage of Fig. \ref{fig2}.

The feature extraction module consists of 6 convolutional layers and 2 fully connected layers. The convolutional layers all use $3\times3$ convolutional kernels. To avoid overfitting, we inserted BatchNormalization layers following each convolution layer, which shrinks the neuron inputs to a normal distribution with a mean of 0 and a variance of 1, rather than a wider random distribution. After every two convolution layers, maximum pooling and dropout layers are added to reduce input data size for the next block. Details of the maximum pooling layers are given in Sect. \ref{Convolution neural network}, and the essence of dropout layers is to randomly discard a certain number of neurons to improve the generalization ability of the model. The output of the fully-connected layer is a feature vector in the form of $128\times1$, which will be passed to the second part for the distance calculation. 

\begin{table}[]
\caption{SC-Net structure in feature extraction process. It consists of three convolutional layers, three pooling layers and two fully connected layers and makes the data from size $64\times64\times3$ to $128\times1$.}\label{tab3}
    \centering
\begin{tabular}{ccc}
\hline
\multicolumn{2}{c}{Layer}        & output-shape \\ \hline
\multicolumn{2}{c}{input layer}  & $64\times64\times3$      \\ \hline
conv-3-32   & BatchNormalization & $64\times64\times32$     \\
conv-3-32   & BatchNormalization & $64\times64\times32$  \\
max-pooling & Dropout 0.5        & $32\times32\times32$    \\ \hline
conv-3-64   & BatchNormalization & $32\times32\times64$     \\
conv-3-64   & BatchNormalization & $32\times32\times64$     \\
ma×-pooling & Dropout 0.5        & $16\times16\times64$     \\ \hline
conv-3-256  & BatchNormalization & $16\times16\times256$    \\
conv-3-256  & BatchNormalization & $16\times16\times256$    \\
max-pooling & Dropout 0.5        & $8\times8\times256$      \\ \hline
\multicolumn{2}{c}{Flatten}      & $16384\times1$      \\
\multicolumn{2}{c}{dense 512}    & $512\times1$        \\
\multicolumn{2}{c}{dense128}     & $128\times1$        \\ \hline
\end{tabular}
\end{table}
 
The Euclidean Distance of two feature vectors is given by 
\begin{equation}
\label{eq8}
D_{v}(x,y)=\sqrt{\sum_{i=1}^{n}(x_{i}-y_{i})^{2}},
\end{equation}
where $x_i$ is the $i$th element of the first feature vector $\vec{x}$, and $y_i$ is the $i$th element of the second feature vector $\vec{y}$. When the $D_{v}$ is less than 0.5, the two images are identified to be sufficiently similar, then classified to be “from the same category”; Otherwise, the images are classified to be “from different categories”. Overall, one needs to train about 9 million parameters in the entire SC-Net model, including the modules of feature extraction and classification.

\subsection{Implementation Details}
\label{Implementation Details}

The hardware system utilized in this study contains: Intel(R) Core(TM) i5-9300H CPU @2.40GHz 2.40 GHz; NVIDIA GeForce RTX 2060 6GB. Software environment comprises python 3.7.3, Keras 2.3.1, NumPy 1.16.2, Matplotlib 3.0.3, OpenCV 3.4.2.16. The total runtime is about 128 hours for ten replicates of 30 experiments. 

\begin{figure*}
\centering
\includegraphics[width=\hsize]{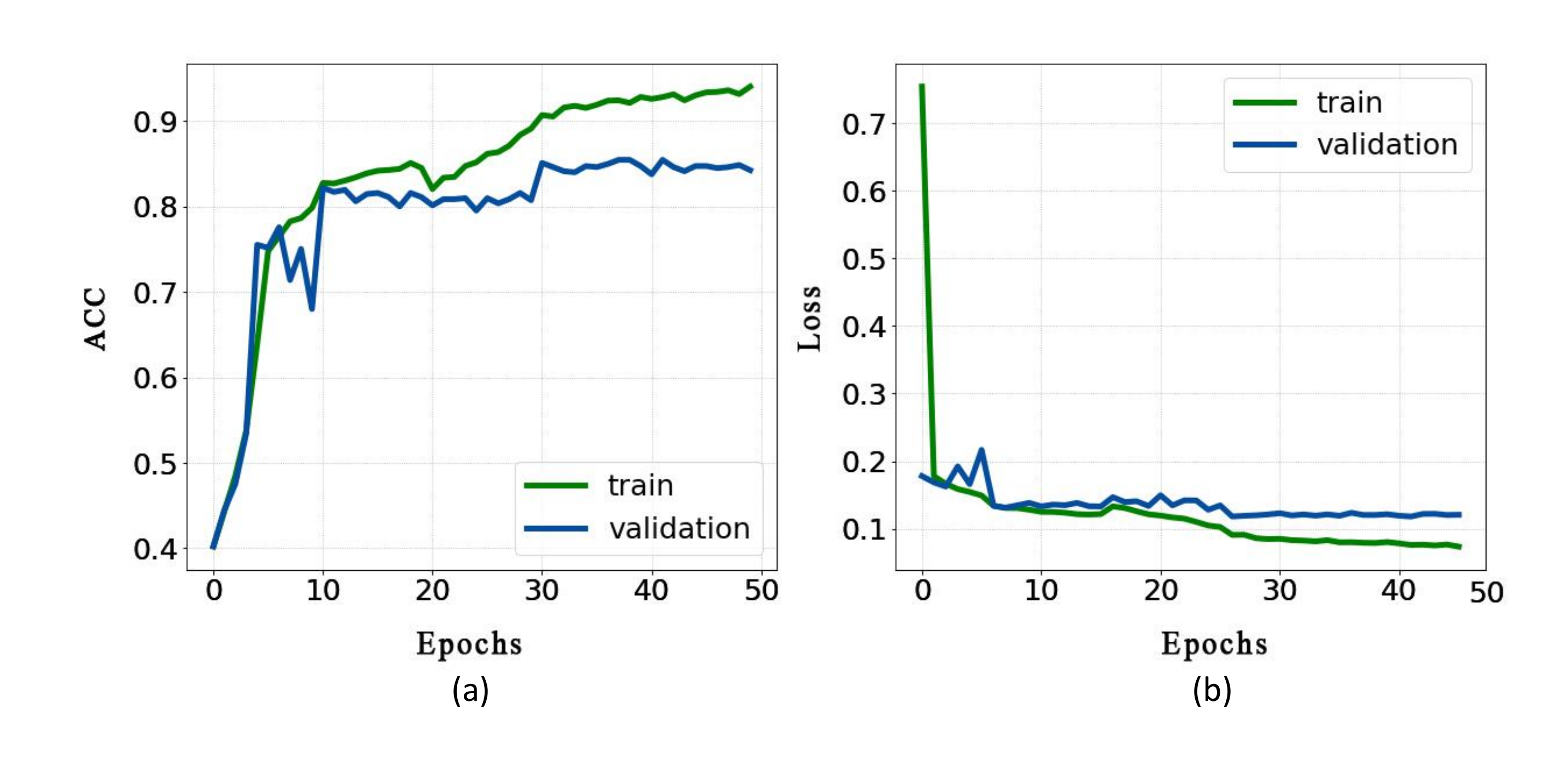}
\caption{The ACC(a) and Loss(b) curves of SC-Net model under the Adam optimizer with 20000 samples in the dataset. In 30 groups of experiments, the iteration times of each experiment was determined according to this figure. We chose the position between 40 to 50 where val-loss gradually flattened with iteration time increasing.}
\label{fig6}
\end{figure*}

In each epoch, the batch-size is set to 32; the loss function is contrastive-loss introduced in Sect. \ref{Siamese network}; the optimizers adopted in the methods based on CNNs are Adam, while we use both Adam and RMS for the SC-Net; the initial learning rate is 0.01, which decreases by ten every ten iterations. Each group experiment was iterated 100 times, and the chosen model was selected according to the ACC and Loss curve. Fig. \ref{fig6} shows ACC and Loss curves of the SC-Net model under the Adam optimizer and 20000 samples in the datasets, we choose a model between 40 and 50 epochs since, at that time, the distance between the validation-loss and training-loss begins to grow, and validation-loss becomes stable, as descried in Fig. \ref{fig6}(a). Likewise, the chosen models based on deep convolutional neural networks are selected between 30 and 40 epochs, and the chosen model based on the SC-Net with RMS optimizer is selected between 50 and 60 epochs. 

\section{Results}
\label{Results}
 
The experiment is performed under six datasets and five models, including three traditional CNNs (AlexNet, VGG\_16, ResNet\_50) and two SC-Net models, i.e., thirty experiments in total. Specifically, the sizes of training sets are 1000, 5000, 10000, 15000, 20000 and 28793 separately. The details of organizing the datasets with different sizes are introduced in Sect. \ref{Datasets}. The five methods are AlexNet, VGG\_16, ResNet\_50, SC-Net Rms, and SC-Net Adam. We adopt accuracy (ACC) as the metric for quantifying the classification performance, which is defined as  
\begin{equation}
\label{eq9}
ACC=\frac{N_{TP}+N_{TN}}{N_{TP}+N_{TN}+N_{FP}+N_{FN}}.
\end{equation}
where $N_{TP}$ stands for the number of true-positives, $N_{TN}$ stands for the number of true-negatives, $N_{FP}$ denotes the number of false-positives, $N_{FN}$ denotes the number of false-negatives.

\begin{figure}[ht!]
\centering
\includegraphics[width=\hsize]{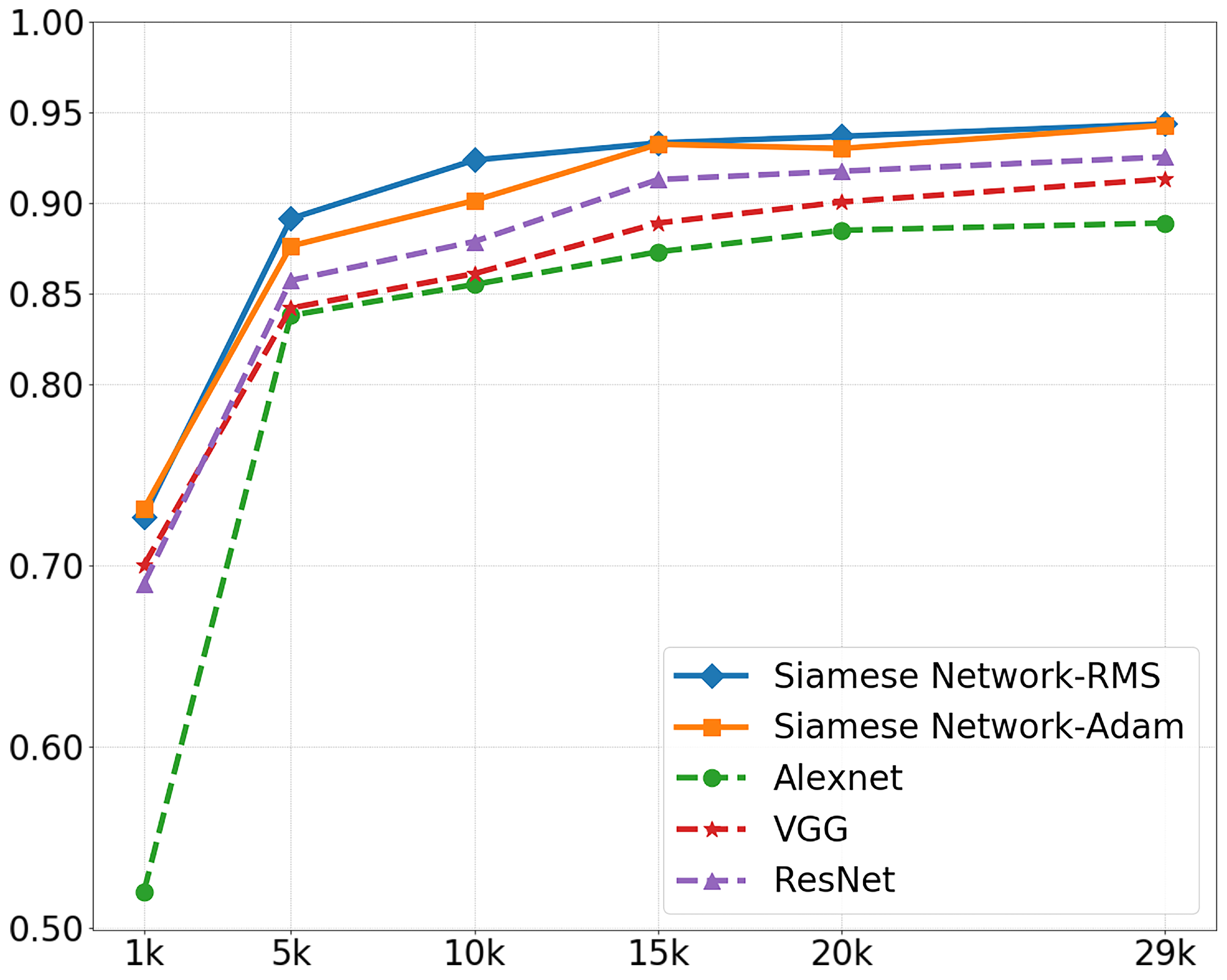}
\caption{Comparison of experimental results of 5 methods of ACC. The vertical axis represents the classification performance, the horizontal axis represents the size of the data set, and the broken lines with different colors represent different methods.}
\label{fig7}
\end{figure}

\begin{figure}[ht!]
\centering
\includegraphics[width=0.7\hsize]{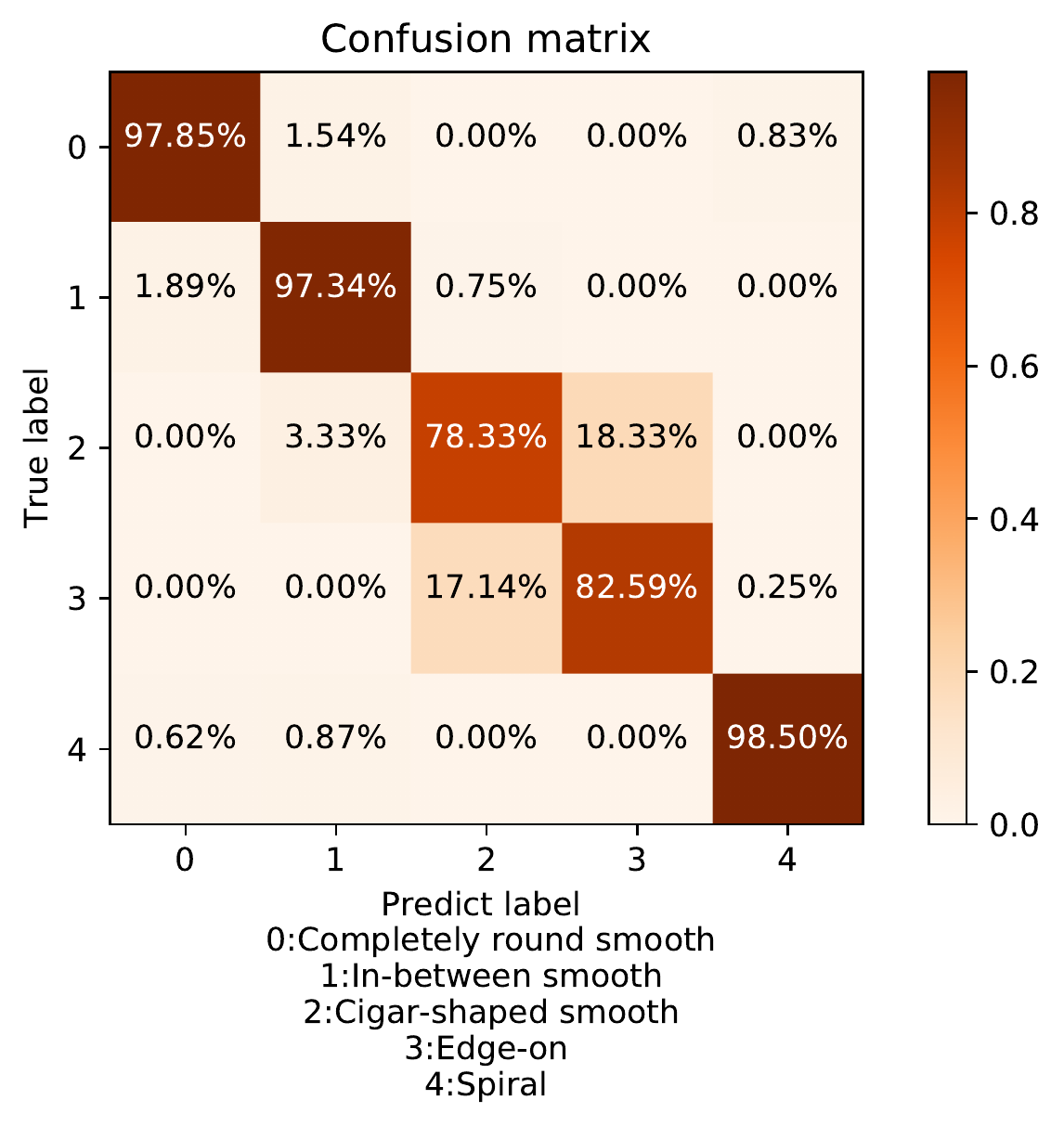}
\caption{Confusion matrix for classification results.The ordinate is the real category of the data, and the ordinate is the category predicted by the model. The diagonal lines represent the percentage of correctly predicted data in each category. The remaining values represent the percentage of predictions that were wrong.}
\label{fig8}
\end{figure}

As is shown in Tab. \ref{tab4} and Fig. \ref{fig7}, the SC-Net model achieves the highest accuracy results with all experimental datasets. The most significant gap is $21\%$ compared to AlexNet when the training set contains 1000 images. Considering the results displayed in Fig. \ref{fig8}, when the training set size is 28793, the ACC of SC-Net is $6\%$ higher than that of AlexNet, one can conclude that less training data leads to more significant excellence of the SC-Net model. This reveals the superiority of the SC-Net model compared to traditional CNNs since the SC-Net model takes paired images and labels \citep{krizhevsky2012,simonyan2014,he2015}, but the CNNs take images and labels directly. Taking paired images and labels enlarges the size of  datasets and magnifies the difference between the images from different morphological categories. In addition, the ACC given by the SC-Net RMS method trained by 10000 images is as high as that given by ResNet\_50 trained by 28793 images. If one plans to acquire a classification ACC of no less than $90\%$, the SC-Net model needs $\sim6300$ images for training, while ResNet\_50 requires $\sim 13000$ images. The reduction of the requirements of training sets enables the useability of the SC-Net model to detect rare objects (such as strong lenses) potentially.

\begin{table*}
\caption{30 groups of experiments, each group of experiments were carried out 10 times, and the median and standard deviation were taken as the final results. The first column represents the size of the datasets, and the next five columns represent the the classification performance of the five methods.}\label{tab4}
    \centering
\begin{tabular}{llllll}\hline
Datasets & AlexNet         & VGG\_16         & ResNet\_50      & \textbf{Sc-Net RMS}      & \textbf{Sc-Net ADAM}     \\ \hline
28793    & $88.89\%\pm0.63\%$ & $91.32\%\pm0.27\%$ & $92.53\%\pm0.43\%$ & $94.36\%\pm0.25\%$ & $94.47\%\pm0.23\%$ \\ \hline
20000    & $88.49\%\pm0.63\%$ & $90.05\%\pm0.78\%$ & $91.75\%\pm0.51\%$ & $93.67\%\pm0.42\%$ & $93.93\%\pm0.30\%$ \\ \hline
15000    & $87.30\%\pm0.69\%$ & $88.89\%\pm0.87\%$ & $91.29\%\pm0.64\%$ & $93.31\%\pm0.56\%$ & $93.23\%\pm0.28\%$ \\ \hline
10000    & $85.50\%\pm0.80\%$ & $86.10\%\pm0.64\%$ & $87.81\%\pm0.60\%$ & $92.36\%\pm0.63\%$ & $91.13\%\pm0.46\%$ \\ \hline
5000     & $83.79\%\pm1.21\%$ & $84.20\%\pm0.89\%$ & $85.72\%\pm0.59\%$ & $89.13\%\pm0.35\%$ & $87.62\%\pm0.37\%$ \\ \hline
1000     & $52.00\%\pm1.11\%$ & $70.00\%\pm0.50\%$ & $69.00\%\pm0.50\%$ & $72.69\%\pm0.62\%$ & $73.12\%\pm0.56\%$ \\ \hline
\end{tabular}
\end{table*}  

We additionally explore the dependence of the classification performance of the SC-Net model on the characteristics of the data. Fig. \ref{fig8} presents the confusion matrix of the SC-Net model, which shows that the SC-Net model can achieve $97.85\%$, $97.34\%$, and $98.50\%$ ACC in the three categories of completely round smooth, in-between smooth, and spiral, since these galaxies have well-identified features. However, the ACC decreases to $78.33\%$ and $82.59\%$ for cigar-shaped and edge-on galaxies because of their similarity in the case of the Point Spread Functions (PSFs) of SDSS. As is mentioned above, the SC-Net model takes paired images and labels to measure their similarity. Thus, its performance may be suppressed when the features of categories are alike. Expectedly, this issue will be less noteworthy when the image quality is improved. For instance, with the data from space-born telescopes, smeared substructures in galaxies will be well resolved, such as bugles, disks, and clumps. Then, the SC-Net model can still separate the cigar-shaped and edge-on galaxies.

\begin{figure}[ht!]
  \begin{adjustbox}
  {addcode=
  {\begin{minipage}{\width}}
    {\caption{The illustration of the distance of feature space between images measured by the SC-Net. Where a, b, c, d, and e stand for that, the images listed along the column are completely round smooth, in-between smooth, cigar-shaped, edge-on, and spiral, respectively. The values describe the similarity between the images in the columns and rows. The smaller the value is, the more similar the two images are. In each image matrix, the first row shows an example of correct classification; the second row shows an example of incorrect classification. Blue boxes denote the category of the data itself, and the fonts in red represent the category predicted by the SC-Net model.
      }
      \label{fig9}
      \end{minipage}},rotate=90,center}
      
      \includegraphics[height=0.47\textheight]{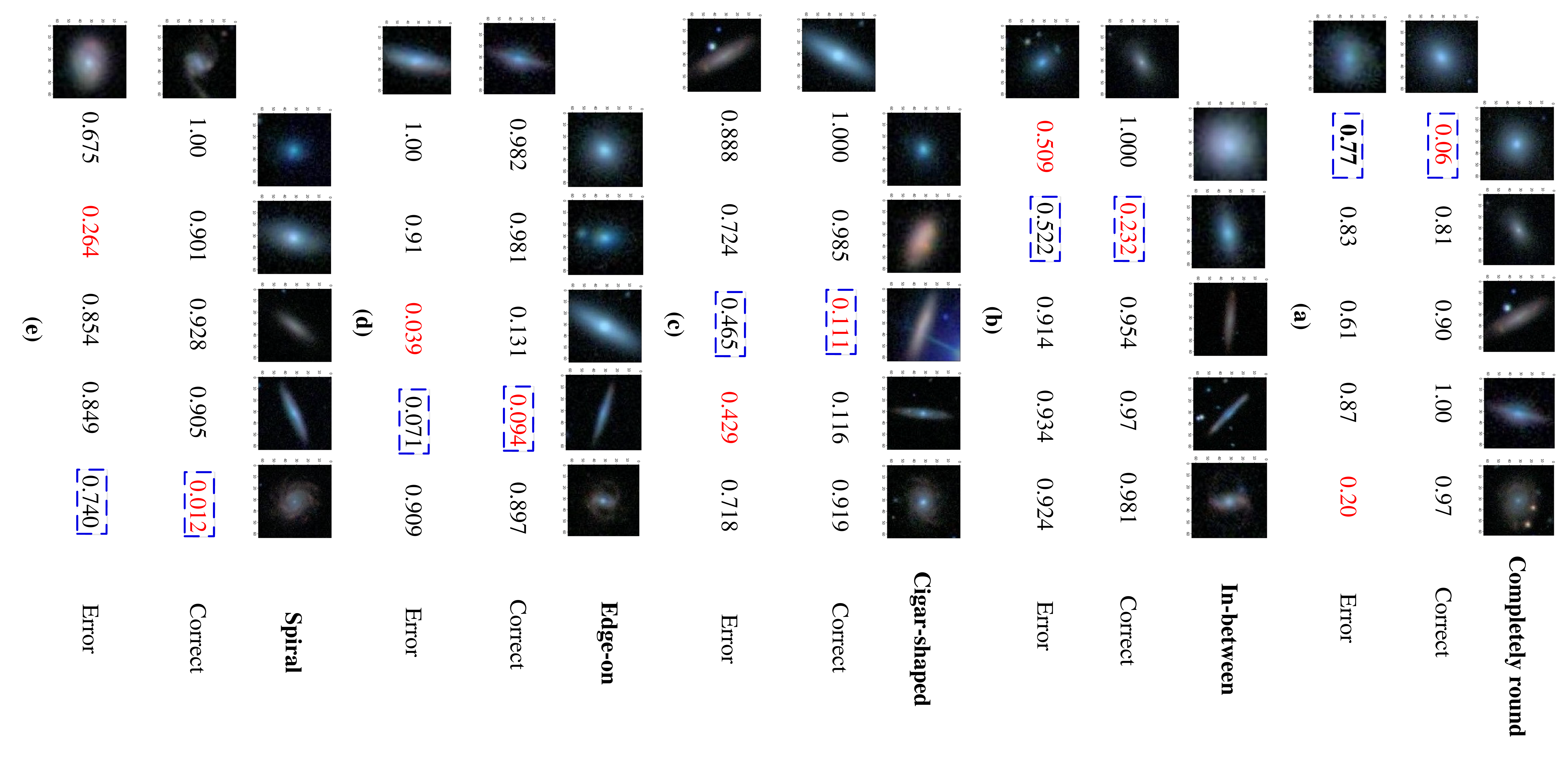}%
  \end{adjustbox}
\end{figure}
 
Moreover, we draw Fig. \ref{fig9} to analyze the correlation between the classification predictions and the distance between testing images and those in different categories in the training sets in feature space. Five panels, a, b, c, d, and e, stand for that: the images listed along the column are completely round smooth, in-between smooth, cigar-shaped, edge-on, and spiral, respectively. The values describe the similarity between the images in the columns and rows. The smaller the value is, the more similar the two images are. In each panel, the first row shows an example of correct classification; the second row shows an example of incorrect classification. Blue boxes denote the ground truth, and fonts in red represent the predicted labels. When the classification is correct in the three categories with apparent features, the similarity between the testing and training images in the corresponding category is quite different from that between the testing images and training images in other categories. For instance, the similarity score in Fig. \ref{fig9}(a) is 0.06 in the case of correct classification, while the other distance of feature space scores are above 0.80. However, for the types of cigar-shaped and edge-on galaxies, the differences are only 0.005 and 0.036, which can be calculated by  0.116 - 0.111 and 0.465 - 0.429, see panel c. These outcomes further prove that similarity between the data points in different categories in the training set considerably influences the accuracy of galaxy morphology classification. Hence, it is critical to organize training sets sensibly to avoid such similarities as much as possible when one plans to adopt the SC-Net to solve their problems. 

\section{Discussion and Conclusion}
\label{Discussion and Conclusion}
Traditional supervised deep learning methods are currently the mainstream for the morphological classification of galaxies, which request a considerable volume of training sets. Suppose it demands simulations to create sufficient training sets, which potentially brings model-dependence problems. Thus, we introduce few-shot learning based on the SC-Net model to avoid these drawbacks. Our results present that Few-shot learning reduces the requirement of the size of training sets and provides an efficient way to extend the coverage of the training sets in latent space, which can be used to avoid the model-dependence problem.

To illustrate the improvements of our method, we conduct comparative experiments between few-shot learning and approaches based on traditional CNNs, such as AlexNet, VGG\_16, and ResNet\_50. The results show that Few-shot learning achieves the highest accuracy in most cases, and the most significant improvement is $21\%$ compared to AlexNet when the training sets contain 1000 images. In addition, to guarantee the accuracy is no less than $90\%$, few-shot learning needs $\sim 6300$ images for training, while ResNet\_50 requires $\sim 13000$ images. The request for fewer training data can avoid simulation as much as possible when constructing training sets, which bypasses the model dependence problem. Further, suppose we design a recursive strategy to enlarge the training set for galaxy morphology classification by starting with a small training set. Then, few-shot learning can start with extensively fewer data points with known labels than those based on traditional CNNs, which remarkably decreases the workload on creating the primary training set, especially for the case of labeling images by human inspection. 

Notably, the performance of few-shots learning is sensitive to the similarity between the images with different labels, though it is still better than that of the methods based on CNNs. For instance, the classification accuracy of completely round smooth, in-between smooth, and spiral are higher than that of cigar-shaped and edge-on. Specifically, the classification accuracy reaches $97.85\%$, $97.34\%$, and $98.50\%$ in completely round smooth, in-between smooth, and spiral. However, in the two categories of cigar-shaped and edge-on, the accuracy is $78.33\%$ and $82.59\%$ separately. It is reasonable since SC-Net adopts the Euclian Distances between images in latent space as the classification metric, while higher similarity leads to shorter distances, which causes miss classification. This issue is primarily due to the limitation of Galaxy Zoo images observed by ground-based telescopes, presenting few small-scale structures because of large PSFs. After all, the difference between cigar-shaped and edge-on is also hard to identify by human inspection. Expectedly, future high-quality images with detailed structures captured by space-born telescopes can improve the classification performance significantly.

In summary, this study presents the feasibility of few-shot learning on galaxy morphology classification, and it has certain advantages compared to traditional CNNs. Next, we plan to apply the method to observations such as DESI Legacy Imaging Surveys \footnote{\url{https://www.legacysurvey.org/}}, the Dark Energy Survey, and the Kilo-Degree Survey \footnote{\url{https://kids.strw.leidenuniv.nl/}}. Also, to further improve the performance of this approach, we will optimize its architecture and hyper-parameters while implementing the above applications. Besides, considering the characteristic of SC-Net, few-shot learning can also be utilized to identify rare objects, e.g., merging galaxies, ring galaxies, and strong lensing systems, which draws our interests intensively as well.

\begin{acknowledgements}
This dataset used in this work is collected from the Galaxy-Zoo-Challenge-Project posted on the Kaggle platform. We acknowledge the science research grants from the China Manned Space Project with NO. CMS-CSST-2021-A01. ZRZ, ZQZ, and YLC are thankful for the funding and technical support from the Jiangsu Key Laboratory of Big Data Security and Intelligent Processing. The authors are also highly grateful for the constructive suggestions given by Han Yang and Yang Wenyu for improving the manuscript.
\end{acknowledgements}
\appendix 
\bibliographystyle{raa}
\bibliography{sample}
\label{lastpage}
\end{document}